\documentstyle[prl,aps,epsf]{revtex}

\twocolumn

\begin{document}
\draft
\title{Weakly Nonlinear Theory of Pattern-Forming Systems with
  Spontaneously Broken Isotropy}

\author{A.~G.~Rossberg, A.~Hertrich, L.~Kramer, W.~Pesch}
\address{Physikalisches Institut der Universit{\"a}t Bayreuth,
  D-95440 Bayreuth, Germany}
\date{\today}
\maketitle
\begin{abstract}
Quasi two-dimensional pattern forming systems with spontaneously
broken isotropy represent a novel symmetry class, that is
experimentally accessible in electroconvection of
homeotropically aligned liquid crystals.  We present a weakly
nonlinear analysis leading to amplitude equations which couple the
short-wavelength patterning mode with the Goldstone mode resulting
from the broken isotropy. The new coefficients in these equations are
calculated from the hydrodynamics. Simulations
exhibit a new type of spatio-temporal chaos at onset. The results are
compared with experiments.
%% < 600 characters
\end{abstract}
\narrowtext
%\section{Introduction}

Ac driven electroconvection  (EC)  in  nematic liquid  crystal   (NLC)
layers is one of the richest systems for  the study of pattern forming
phenomena    \cite{rw,annrev}.   We consider   the  typical thin-layer
geometry with the (slightly  conducting) NLC sandwiched between  glass
plates coated with transparent electrodes.  An ac voltage U is applied
across the  electrodes. By appropriate surface  treatment of the glass
plates the director ${{\bf \hat n}}$ (the preferred orientation of the
NLC molecules) can be fixed at the boundaries.  In particular the case
of planar alignment (${{\bf  \hat n}}$ in the plane  of the layer) has
been studied  intensely.  At  onset one may   then have normal  rolls,
where   the wavevector is  parallel   to the undistorted director,  or
oblique ones.

The case of homeotropic surface anchoring  (${{\bf \hat n}}$ aligned
perpendicularly to the boundaries) where the system is isotropic in the 
plane of the layer (= $x$-$y$ plane) offers novel possibilities. 
Then, in the traditional
EC materials with negative dielectric anisotropy, the voltage applied
across the layer will first turn the director away from the layer normal
(bend Fre\'edericksz transition) leading to a quasi-planar director 
configuration (see e.g. \cite{degennes}).
The spontaneously chosen direction of the bend (i.e. direction of the
projection of ${{\bf \hat n}}$ on the $x$-$y$ plane) will be denoted by 
${\bf \hat c}={\bf \hat c}({\bf r},t)$ ($\,|{\bf \hat c}|\!=\!1\,$).
Further increase of the voltage will eventually generate EC in  
close analogy
to the planar case \cite{hdpk};
in fact nucleation to normal, oblique and traveling rolls has
been observed \cite{ribure}. The notable difference to the planar case
is that the preferred axis (the in-plane director ${{\bf \hat c}}$) is
degenerate and  not
fixed externally (neglecting unavoidable inhomogeneities and in the
absence of a planar magnetic field).  Then oblique rolls are not
expected to lead to a stable pattern because they will in general
exert a torque on ${{\bf \hat c}}$ which cannot be compensated.  Even normal
rolls, where a torque is absent because of symmetry,
can be unstable because transverse modulations can be enhanced by the torque.

Here we will investigate the situation by setting up a novel weakly 
nonlinear description that incorporates the critical convection mode
together with the Goldstone mode resulting from the spontaneous
breaking of the $O(2)$ symmetry by the bend Fre\'edericksz transition.
The general form of the equations is derived from symmetry 
considerations.
Analyzing the stability of rolls indicates that one may expect
spatio-temporal chaos (STC) at onset under very general conditions.
In fact this is probably the first experimentally accessible example
of a nonrotating system with a direct transition to STC in a stationary
(non-Hopf) bifurcation~\cite{ribure,kai}. The new coefficients of the  
equations are then
calculated from the underlying standard hydrodynamic theory making 
quantitative
comparison with experiments on MBBA
({\sl N}-({\sl p}-methoxybenzylidene)-{\sl p}-butylaniline)
without adjustable parameters in principle possible.
We end up by showing results of simulations.

%\section{Construction of the Equations}
\label{sec:equations}
First we consider the situation where the orientation of ${{\bf \hat c}}$
differs only by a small angle $\varphi$ from an overall direction  
that we
choose as the $x$ axis
(then ${\bf \hat c} = (cos \varphi, sin \varphi,0) \approx (1,\varphi,0)$). 
For simplicity we present the analysis for the normal-roll case, but
it was done also for oblique rolls and some results will be given
for that case.
Let ${\bf q}_c=(q_c,0)$, be the wavevector of the most unstable convection
mode when $\varphi=0$.
Near threshold the physical state can be written in the form
\begin{eqnarray}
  \label{physical-form}
  {\bf V}({\bf r},z,t)=& 
  \bigl( & {\bf V}^{(1)}(z,t) A({\bf r},t) e^{i {\bf q}_c \cdot  
{\bf r}}+
    \text{c.c.} \bigr) \\
  &+& {\bf V}_G(z,t) \varphi({\bf r},t)+h.o.t.,  \nonumber
\end{eqnarray}
where ${\bf V}$ is a formal vector consisting of the three components of the 
velocity, the director ${\bf \hat n}$, and the charge density.
$A=\text{\it const.}$ and $\varphi=0$ yields the critical mode  at the
threshold $0=\varepsilon = (U^2-U_c^2)/U_c^2$.  The reduced description
in terms of the modulation amplitudes $A$ and $\varphi$ has to
respect the symmetries of the underlying hydrodynamic equations and the
Fre\'edericksz state. Thus we require invariance of the equations
under translation (${\bf r} \to {\bf r} + {\bf r}_0 ,\, A \to A e^{-i
  {\bf q}_c \cdot {\bf r}_0}$), rotation ($ \varphi \to \varphi + \delta,\,
{\bf r} \to {\bf r} + \delta {\bf \hat z} \times {\bf r},\, \partial_x \to
\partial_x - \delta \partial_y,\, \partial_y \to \partial_y + \delta
\partial_x,\, A \to A e^{i \delta {\bf \hat z} \times {\bf q}_c \cdot \vec
  r}$), three dimensional inversion ($ x \to -x,\, y \to -y,\, A \to
A^* $), reflection at
the $x$ axis ($ y \to -y,\, \varphi \to -\varphi$),
and time-translation ($ t \to t + t_0$).
We consider the regime near onset ($\varepsilon,|A| \ll 1$) where slow
spatial and temporal variations ($\partial_x,\partial_y,\partial_t  
\ll 1$)
can be assumed. However, similar to~\cite{ber}
no specific scaling relations are assumed at this point.

By   retaining only the    leading  terms allowed by
symmetries   the  equations describing  the   evolution  of   $A$ and
$\varphi$ can be determined up to prefactors:
 
\begin{eqnarray}
  \label{plain-NW-normal-A}
  \tau \partial_t A &=&
    \Big [\varepsilon+\xi_{xx}^2  
\partial_x^2+\xi_{yy}^2(\partial_y-i q_c \varphi)^2
  \\
  &-&g|A|^2 + i \beta_y \varphi_{,y}\;
  \Big ]\, A,
  \nonumber
\end{eqnarray}
\begin{eqnarray}
  \label{plain-NW-normal-phi}
  \tilde \gamma_1 \partial_t \varphi &=&
    K_1 \partial_{y}^2 \varphi+K_3 \partial_{x}^2 \varphi
    - \tilde \chi_a H^2 \varphi
  \\
  &+& {\Gamma \over 4}
  (-i q_c A^* (\partial_y - i q_c \varphi) A + \text{c.c.}).
  \nonumber
\end{eqnarray}
In (\ref{plain-NW-normal-A}) and in similar equations the derivative
operators $\partial_x$, $\partial_y$ operate on $A$ only, while
$\varphi_{,y}$ is the derivative of $\varphi$ in the $y$
direction.  All coefficients are real and we will assume all but
$\beta_y,\Gamma$ to be positive. 
The coefficients $\beta_y$ and $\Gamma$ describe the novel effects.
The last expression in
Eq.~(\ref{plain-NW-normal-phi}) describes the torque
exerted by the roll pattern on
${\bf \hat c}$.
An additional term $-\tilde \chi_a H^2 \varphi$
is included in Eq.~(\ref{plain-NW-normal-phi})
to allow a continuous transition from the rotationally invariant
case to the case of fixed ${{\bf \hat c}}$, as for planarly aligned cells.
Physically $H$ corresponds to a stabilizing magnetic field parallel to 
the $x$-axis.
In the special case $\beta_y=-\xi_{yy}^2 q_c$ the angle $\varphi$ would
play the role of a gauge field and Eq. (\ref{plain-NW-normal-A}) could
be derived from a potential. If also $\Gamma>0$, both equations can be
derived from a Lyapunov functional. However, in our non-equilibrium
system there is no reason to assume existence of a potential.
For MBBA we find $\Gamma < 0$, which implies a ``repulsive torque''  
between
the direction of the local wavevector and ${\bf \hat c}$.
Then at $H=0$ all ordered roll solutions are in fact unstable (see  
below).

For $H=0$
Eqs.~(\ref{plain-NW-normal-A},\ref{plain-NW-normal-phi}) become
uniform in $\varepsilon$
when the scaling  $\partial_x,\partial_y,A,\varphi \sim  
\varepsilon^{1/2}$
and $\partial_t \sim \varepsilon$ is adopted.
This is a major difference to similarly
looking coupled amplitude equations describing mean-flow effects in  
isotropic
Rayleigh-B\'enard convection~\cite{ber,szmf,dp}.
Since $\varphi$ is also scaled, this is not applicable to
Eqs.~(\ref{ri-plain-NW-normal-A},\ref{ri-plain-NW-normal-phi}) below
because there $\varphi$ is $2\pi$ periodic.

After rescaling~(\ref{plain-NW-normal-A},\ref{plain-NW-normal-phi})
one arrives at the normal form
\begin{eqnarray}
  \label{scaled-NW-normal-A}
  \check \tau \partial_t \check A &=&
    \Big [1+ \partial_x^2+(\partial_y-i \check \varphi)^2
  \\
  &-&|\check A|^2 + i \check \beta_y \check \varphi_{,y}\;
    \Big ]\, \check A,
  \nonumber
\end{eqnarray}
\begin{eqnarray}
  \label{scaled-NW-normal-phi}
  \partial_t \check \varphi &=&
    \partial_{y}^2 \check \varphi+\check K_3 \partial_{x}^2 \check  
\varphi
    - \varepsilon^{-1} \check H^2 \check \varphi
  \\
  &+& \check \Gamma \;
  (-i \check A^* (\partial_y - i \check \varphi) \check A + \text{c.c.}),
  \nonumber
\end{eqnarray}
where
$A=(\varepsilon/g)^{1/2}\check A$,
$x=\xi_{xx} \varepsilon^{-1/2} \check x$,
$y=\xi_{yy} \varepsilon^{-1/2} \check y$,
$\varphi=\varepsilon^{1/2} \check \varphi/(\xi_{yy} q_c)$,
$t=\tilde \gamma_1 \xi_{yy}^2 \check t/(\varepsilon K_1)$,
$\check \beta_y=\beta_y/(\xi_{yy}^2 q_c)$,
$\check K_3 = K_3 \xi_{yy}^2/(K_1 \xi_{xx}^2)$,
$\check H^2=\chi_a \xi_{yy}^2 H^2 /K_1$,
$\check \Gamma=q_c^2 \xi_{yy}^2 \Gamma/(4 g K_1)$,
$\check \tau=K_1 \tau/(\gamma_1 \xi_{yy}^2)$.
The caret $\check {\text \ }$
has been suppressed
on the variables $x$, $y$, $t$ in 
Eqs.~(\ref{scaled-NW-normal-A},\ref{scaled-NW-normal-phi}).

The stationary, uniform solutions of these
equations are $\check A=\check A_0 e^{i(Q x + P y)}$, $\check
\varphi=\check \varphi_0=2 \check \Gamma \check A_0^2 P/(\check  
H^2/\varepsilon+2 \check \Gamma \check A_0^2)$, $A_0^2=1-(P-\check  
\varphi_0)^2-Q^2$.
We now present the main results of the stability analysis of
these roll solutions. For $\check \Gamma<0$ and $H=0$ rolls are always
unstable at onset, and one expects a (direct) transition to
spatio-temporal chaos. For $\check H \neq 0$ and $\check \beta_y<0$ one has a
 kind of
Eckhaus instability associated with a finite angle between the director
and the local wavevector of the rolls, whereas for $\check \beta_y>0$
an instability 
induced by director modulations via the $\check \beta_y$-term
in~(\ref{scaled-NW-normal-A}) is dominating.

Perturbations can in general be cast in the following form:
\begin{eqnarray}
  \label{pertub-form}
   \delta \check A&=&(A^+ e^{s t +i (k_x x
  +k_y y)} + A^- e^{s^* t -i (k_x x +k_y y)}) e^{i(Q x + P y)},\nonumber\\
  \delta \check \varphi&=&(\tilde \varphi e^{s t + i(k_x x + k_y y)} +
  \text{c.c.}).
\end{eqnarray}
The case $\check \varphi_0=0=P$ is particularly simple.
Then the constant-amplitude solutions are stable if
$\check H^2/\varepsilon+2 \check \Gamma (1-Q^2)>0$,
$\check H^2/\varepsilon+2 \check \Gamma (1-Q^2)(1+\check \beta_y)>0$ and
$Q^2<1/3$ (Eckhaus criterion).
For $\check H=0$ stable rolls require $\check \Gamma>0$,
$1+\check \beta_y>0$ and $Q^2<1/3$. Otherwise one has
stability if $P \check \Gamma/\check \varphi_0 >0 $,
$\check \beta_y \check \varphi_0/P > -1$ and $\det M_{ij}>0$, where
\begin{eqnarray}
  \label{crit_mat}
      M_{xx}&=&1-(P-\check \varphi_0)^2(1+2 \check \varphi_0/P) - 3 Q^2,
       \nonumber\\
      M_{xy}&=&M_{yx}=(P-\check \varphi_0) Q
        ((1-\check \beta_y) \check \varphi_0/P-2) ,\\
      M_{yy}&=&(1+\check \beta_y \check \varphi_0/P)
        (1-3(P-\check \varphi_0)^2-Q^2). \nonumber
\end{eqnarray}
The direction of $(k_x, k_y)_{\text{crit}}$
at the marginal point $\det M_{ij}=0$ is
given by the null space of $M_{ij}$.

At the generalized Eckhaus boundary $(P-\check \varphi_0)^2+Q^2=1/3$
known for planarly aligned EC~\cite{annrev}, $\det M_{ij}\sim -(\check
\varphi_0/P)^2$ is already negative. Thus, as expected, the
additional degree of freedom $\varphi$ leads to a destabilization of the
system.

Expanding for small $P$ and $Q$ we find the uniform solution to be
stable for $\check H^2 >  \check H^2_{\text{crit}}$ where
\begin{equation}
  \label{Hcrit2}
  {\check H^2_{\text{crit}} \over \varepsilon}=
      -2 \check \Gamma \check A_0^2 \left(
        1+3^{1/2}|P|+3 P^2
        + {\text{h.o.t.}} \right)
\end{equation}
if $\beta_y<0$ and
\begin{equation}
  \label{Hcrit}
  {\check H^2_{\text{crit}} \over \varepsilon}=
      -2 \check \Gamma \check A_0^2 \left(
        1+\check \beta_y
        +{(1+\check \beta_y)^4 \over \check \beta_y^3} P^2 Q^2
        + {\text{h.o.t.}} \right)
\end{equation}
if $\beta_y>0$.
Since in the latter case the r.h.s.\ is
dominated by the quadratic dependence of $\check A_0^2$ on $P$, $Q$, 
small but finite $P$, $Q$ are more stable then $P,Q=0$.

We expect that an experimental verification of the stability analysis
is possible. Note that the value of $P$ can be adjusted
by turning the magnetic field with respect to the probe after a  
stable roll
pattern has been established.

The analogous equations for the amplitudes $A_1$, $A_2$ of oblique
rolls (zig and zag) have additional terms and the expression for the
torque on ${\bf \hat c}$ acquires the form
${\Gamma q_c p_c} (|A_1|^2 - |A_2|^2)/2$.
If the crossed-roll solution $|A_1|=|A_2|$ is preferred over simple  
rolls
(e.g. $A_1 \neq 0$, $A_2=0$) the
situation is presumably analogous to that of normal rolls (no  
torque on the
director for the unmodulated solution).
In the opposite case (as for MBBA) the resulting equations have no
nontrivial time-independent roll solution. In that case $\varepsilon$
cannot be scaled out completely and one gets different scaling ranges.

%\section{Algorithm to Calculate the Coefficients}
\label{sec:coefficients}

Next we describe the calculation of the coupling
coefficients $\beta_y$ and $\Gamma$
from the standard hydrodynamic equations for NLCs \cite{chl,bzk}.
The other coefficients are obtained following standard procedures
(see e.g.\ \cite{bzk,manbook}).
As usual the set of equations
is decomposed into a sum of
operators operating on the state vectors ${\bf V}={\bf V}({\bf r},z,t)$. Each
operator is linear in each of its arguments:
\begin{equation}
  \label{multilinear-decomposition}
  0=O({\bf V})=-B \partial_t {\bf V}+L {\bf V}+N_2({\bf V}|{\bf V})+\dots
\end{equation}
The operator $O({\bf V})$ depends on $\varepsilon$ in two ways: by the
explicit action of the electric field and the field dependence
of the Fre{\'e}dericksz state, which is the basic state in this
formalism.

Two components can be distinguished in the kernel of $L$: the
convective mode ${\bf V}^{(1)}$ at wavevectors ${\bf q}=\pm {\bf q}_c$
and the Goldstone mode ${\bf V}_G$ at ${\bf q}=0$ \cite{hdpk}
normalized such that it corresponds to
$\delta n_y = \partial {{\bf \hat c}}/\partial \varphi|_{\varphi=0}$.
Near onset these two modes govern the dynamics described in
Eqs.~(\ref{plain-NW-normal-A},\ref{plain-NW-normal-phi}).
We will also need the linear eigenmodes ${\bf V}^{(1)}(p)$, ${\bf V}_G(p)$ of
(\ref{multilinear-decomposition})
into which ${\bf V}^{(1)}$ and ${\bf V}_G$ develop for transverse modulations,
i.e. at ${\bf q}=\pm(q_c,p)$ and $\pm(0,p)$ respectively.
The coefficients are obtained in
 some analogy to the treatment of mean-flow effects \cite{ber,szmf,dp}
and of the concentration field in binary fluids \cite{riebin}.

Let $\langle {\bf V}_1,{\bf V}_2 \rangle$ be a scalar product
and ${{\bf V}^{(1)}}^+(p)$ and ${{\bf V}_G}^+$ be the 
adjoint eigenvectors of $L$
corresponding to ${\bf V}^{(1)}(p)$ and ${\bf V}_G$. 
The coefficient $\beta_y$ in 
Eq.~(\ref{plain-NW-normal-A}) turns out to be
\begin{eqnarray}
  \label{cy}
  \beta_y\! = \! && -
  \langle{{\bf V}^{(1)}}^+,(\partial_\varepsilon L) {\bf V}^{(1)}\rangle^{-1}
  {d \over d p}
  \biggl \langle {{\bf V}^{(1)}}^+(p),
  \\
  && \left( N_2({\bf V}^{(1)}|{\bf V}_G(p))+N_2({\bf V}_G(p)
  |{\bf V}^{(1)})\right) \biggr \rangle
  \biggr|_{p = 0 \atop \varepsilon=0}.
  \nonumber
\end{eqnarray}
The coefficient of the hydrodynamic angular momentum in
(\ref{plain-NW-normal-phi}) is
\begin{equation}
  \label{Gamma-normal}
  {\Gamma q_c \over 2} = C {d \over dp}{\langle {{\bf V}_G}^+
    ,N_2({\bf V}(p)|{\bf V}(p))\rangle}
  \Big |_{p=0} ,
\end{equation}
where $C$ is chosen such, that $\tilde \gamma_1=1$.

%\section{Comparison with experiments}
\label{sec:experiments}

In order to explore the attractor in the roll-unstable case and 
to compare the long-time dynamics with experiments
the assumption that $\varphi$ be small must be relaxed
(at least at $H=0$). One 
in fact only needs the
condition, that the local wavevector differs little from a (local)
critical wavevector $q_c {{\bf \hat c}}$.  Formally this can  
be achieved by
reintroducing the fast variation ($\tilde A=A e^{i {\bf q}_c \cdot  
{\bf r}}$)
and rewriting derivatives and the magnetic torque in
rotation-invariant form 
(we write ${\bf \hat c}_\perp={\bf \hat z} \times {\bf \hat c}$):
\begin{eqnarray}
  \label{ri-plain-NW-normal-A}
  \tau \partial_t \tilde A &=& 
  \Big [\varepsilon+\xi_{xx}^2 ({\bf \hat c} \cdot \partial_{{\bf r}}-iq_c)^2
  +\xi_{yy}^2({\bf \hat c}_\perp \cdot \partial_{{\bf r}})^2  
  \\
  &-&g|\tilde A|^2 
  + i \beta_y {\bf \hat c}_\perp \cdot \nabla \varphi
  \Big ]\, \tilde A,
  \nonumber
\end{eqnarray}
\begin{eqnarray}
  \label{ri-plain-NW-normal-phi}
  \gamma_1 \partial_t \varphi &=&
  K_3 {\bf \hat c}_\perp \cdot \nabla^2 {{\bf \hat c}}
  + (K_1-K_3) {\bf \hat c}_\perp \cdot \nabla (\nabla \cdot {{\bf \hat c}})
  \\
  &+& \chi_a ({\bf \hat c} \cdot {\bf H})({\bf \hat c}_\perp \cdot {\bf H})
  + {\Gamma \over 4} 
  (-i q_c \tilde A^* ({\bf \hat c}_\perp \cdot \partial_{{\bf r}}) \tilde A\! 
      +\!\text{c.c.}).
  \nonumber    
\end{eqnarray}
These equations have been used for numerical simulations.

\begin{figure}[htbp]
  \begin{center}
    \leavevmode
    \vbox{
      \hbox{
      \epsfxsize 1.5 in
      \epsfysize 1.5 in
      \epsfbox{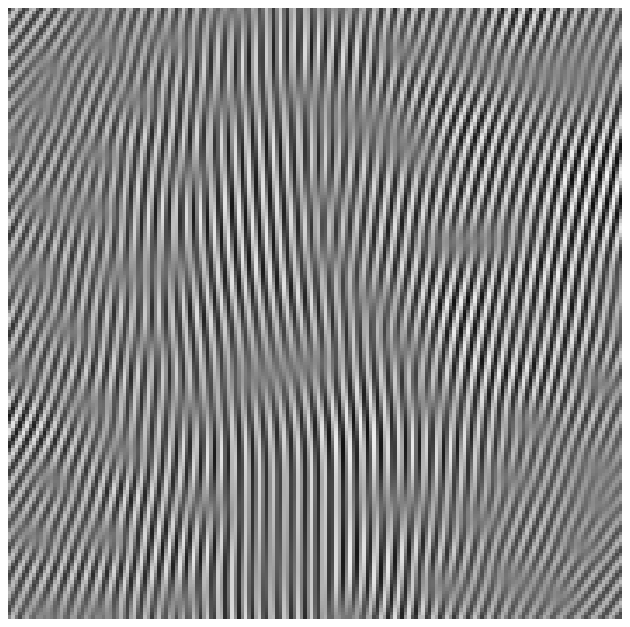}
      \epsfxsize 1.5 in
      \epsfysize 1.5 in
      \epsfbox{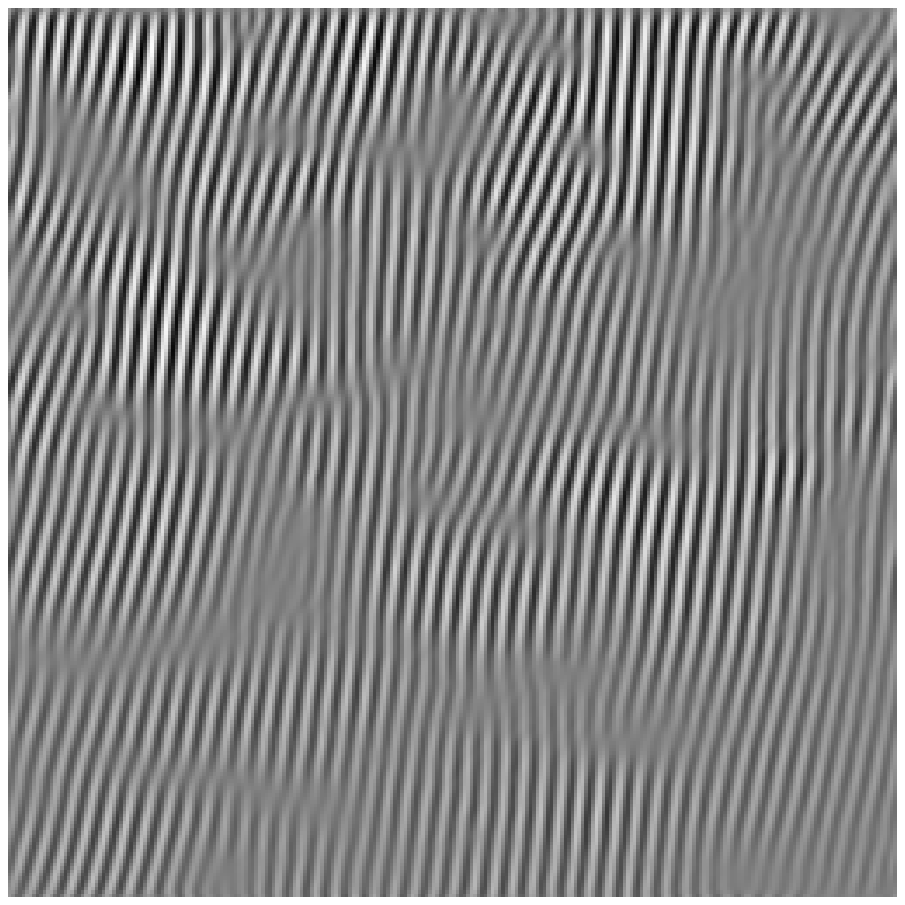}
        }
      \hbox{
      \epsfxsize 1.5 in
      \epsfysize 1.5 in
      \epsfbox{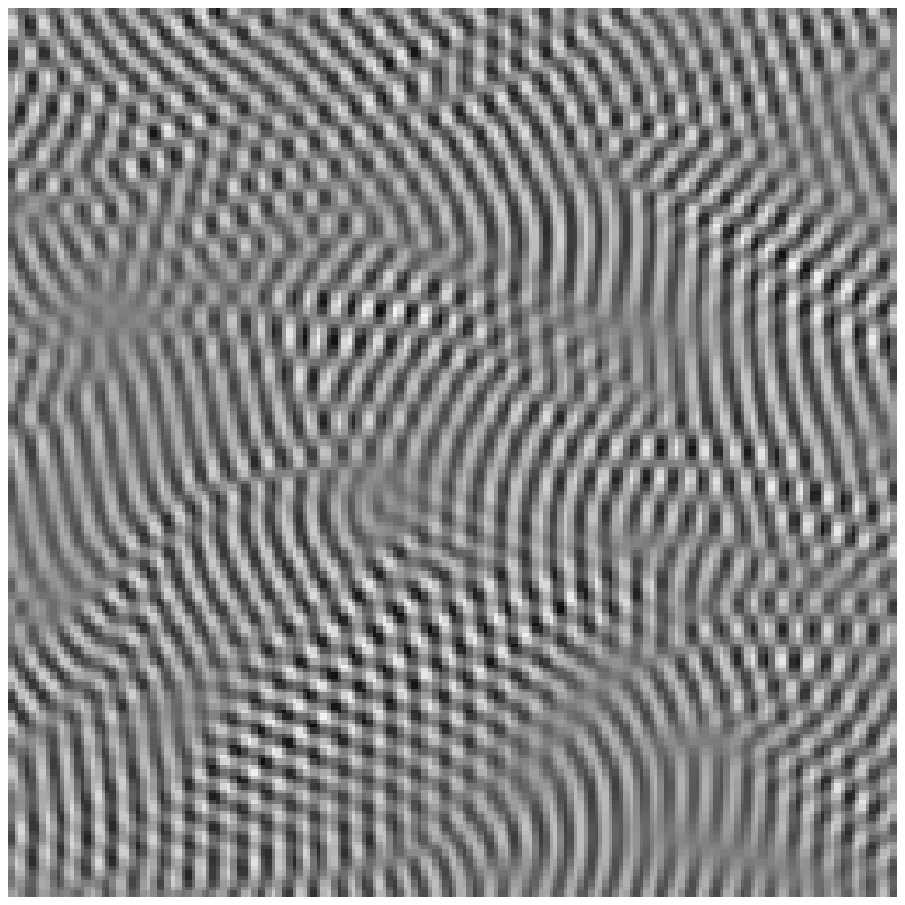}
      \epsfxsize 1.5 in
      \epsfysize 1.5 in
      \epsfbox{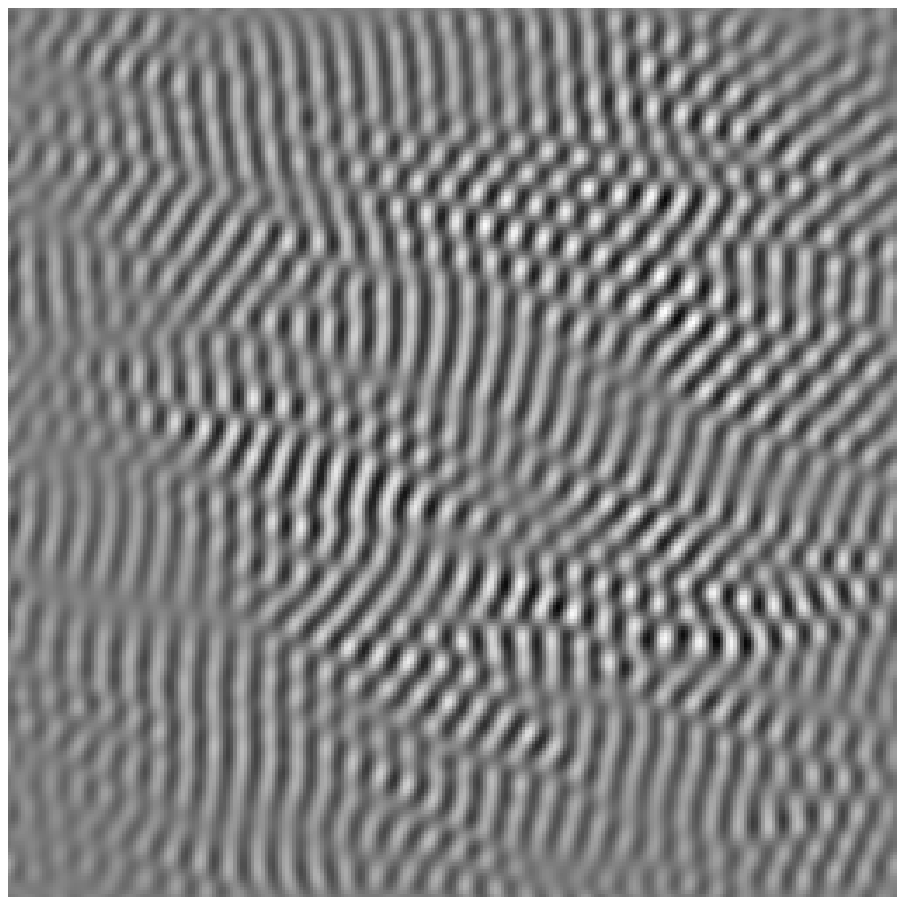}
        }
      }
    \caption{Simulations of Eqs.  
(\protect\ref{ri-plain-NW-normal-A},\protect\ref{ri-plain-NW-normal-phi})
     (left) vs. experiment \protect\cite{ribure1} (right) for  
normal (top) and
        oblique (bottom) rolls (see text).}
    \label{fig:vorzeige}
  \end{center}
\end{figure}

In Fig.~\ref{fig:vorzeige} we compare snapshots of $\psi=\tilde A + c.c.$
from our
simulations (left side) with experiments (right side) from
\cite{ribure,kai} ($\varepsilon=0.02$, $H=0$).
Material parameters as in \cite{hdpk} were used to
calculate the coefficients of 
(\ref{ri-plain-NW-normal-A},\ref{ri-plain-NW-normal-phi}). 
For normal rolls (upper pair), at the
experimental driving frequency $\omega \tau_0=1.75$
($\tau_0=\epsilon_0 \epsilon_\bot /~\sigma_\bot$ is the charge
relaxation time) the calculated coefficients are: $\tau=0.0292$,
$q_c=2.71$, $\xi_{xx}^2=0.434$, $\xi_{yy}^2=0.225$, $\beta_y=-0.60$,
$K_1=1.01$, $K_3=1.28$, $g=4.37$, $\Gamma=-12.8$, $\chi_a=1$.
(We measure length in units of
$d/\pi$ with $d$ the thickness of the fluid layer, time in director
relaxation times $\tau_d=\gamma_1 d^2 / k_{11} \pi^2$ and magnetic
fields in Fre{\'e}dericksz fields $H_f^2 = k_{11} \pi^2 / \chi_a d^2$.)
For oblique rolls (at $\omega \tau_0=0.73$, lower pair) we proceeded
analogously. In order to obtain a good fit to the wavelength the
thickness $d=28.5\, {\rm \mu m}$ was used instead of the nominal value
$d=20\,{\rm \mu m}$ given in \cite{rith} ($d$ was not measured
directly). This only rescales length.

The patterns in experiments and simulations look very similar.
Though in the normal roll regime some defects appear, the rolls are
locally aligned along some main direction. In the simulations we find
the pattern to change on a time scale $t_c/\epsilon =0.8$ 
($t_c$ is defined by $\left<\psi(\vec
  r,t-t_0) \psi({\bf r},t_0)\right>_{{\bf r},t_0} \sim \exp(-|t|/t_c)$)
while the experimental state appeared to be time independent.
Presumably some pinning of the pattern in the experiments
is responsible for this disagreement.
The oblique roll regime is dominated by a superposition of zig and
zag. Again the preferred axis changes only over large distances.
A persistent time dependence was observed in simulations
($t_0/\epsilon=0.5$) and experiments. 
Unfortunately the experiments are still preliminary so far. 
A systematic experimental investigation is under way~\cite{tobu}.

Equations (\ref{scaled-NW-normal-A},\ref{scaled-NW-normal-phi})
represent normal-form equations for quasi-2D pattern-forming systems
with a novel kind of symmetry, and should thus be of general interest.
Other realizations might be found in convection instabilities in
smectic-C liquid crystals, where one has a ${\bf \hat c}$-director from the
beginning on, and in Rayleigh-B\'enard convection of homeotropically
aligned NLCs with an additional electric field. In this case the
fields that drive the Fre{\'e}dericksz transition and the convection
instability, respectively, can be varied independently, which  
should allow
to access a large parameter range of Eqs.
(\ref{scaled-NW-normal-A},\ref{scaled-NW-normal-phi}).

For $H=0$ and $\Gamma < 0$ Eqs.  
(\ref{ri-plain-NW-normal-A},\ref{ri-plain-NW-normal-phi}) describe
STC at onset.
There is some analogy to a recent investigation of a 1d model for
seismic waves, where a spontaneous symmetry breaking is also
important~\cite{trib}. 
Other examples of STC at onset are the
K\"uppers-Lortz instability (in rotating Rayleigh-B{\'e}nard
convection \cite{kuelo,huecah}) and systems
undergoing a Hopf bifurcation. Here one has the Benjamin-Feir  
destabilization
mechanism (see e.g. \cite{croho} and as a recent example
 \cite{dca}) and the so-called dispersive chaos where a quantitative 
description by a simple
Ginzburg-Landau equation should be possible \cite{kolodner}.
The origin of chaotic behavior is of course very different
in the various systems and their detailed comparison appears fruitful.
Work on the characterization of the complex spatio-temporal states is
in progress, with particular emphasis on the transition to order under
the influence of the stabilizing magnetic field.

For ordinary stationary bifurcations (and Hopf bifurcations in the
Benjamin-Feir stable range) the onset of STC is presumably always
at finite $\varepsilon$ (possibly quite small), at least as long as 
effects from lateral boundaries play no role.
Examples are the usual case of EC with planar alignment \cite{kp},
besides our system in the presence of a stabilizing magnet field.

We wish to thank A.~Buka and H.~Richter for discussions and providing
us with unpublished materials.  This work was supported by Deutsche
Forschungsgemeinschaft (Graduierten-Kolleg ``Nichtlineare
Spektroskopie und Dynamik'') and Volkswagen Stiftung.

\bibliographystyle{prsty}

%\onecolumn
%\widetext

\end{document}